\documentclass [twocolumn] {article}

\usepackage{graphics}

\begin{document}

\title{\textbf {Was the Whirlpool (M51) a Spiral Galaxy before the Encounter with the Companion NGC 5195?}}

\author{\small F. Natali, L. Di Fino, G. Natali \\ \and \footnotesize CNR-IAS, Area Ricerca Tor Vergata, Via del Fosso del Cavaliere, 1 - 00133 Roma , Italy; email: natali@saturn.ias.rm.cnr.it }

\date{ }
          
\maketitle 

   \begin{abstract}
Using the B-I color index in the radial photometry of M51, applied to defined azimuthal sectors, we reveal asymmetries in the shape of the density pattern of the galactic arms. The anomalous shape is present only in the side close to the companion and not in the far side. Two more non-interacting check-spirals do not show similar behaviour. The anomalous sawtooth-like shape is similar to that produced by a progressive wave propagating in a preexisting density pattern. In this hypothesis we have to conclude that the galaxy M51 could be a spiral galaxy formerly, i.e. before the encounter with the companion.
 
   \end{abstract}

%________________________________________________________________

\section{Introduction}
Among spiral galaxies, one of the most extensively observed object is undoubtedly the system of M51 (NGC 5194) and its companion NGC 5195. The first galaxy in which spiral structure was observed  since 1845, this system has been studied in every e.m. wavelengths with ground based telescopes, balloon borne and space instrumentation. Several studies in morphology, spectroscopy, photometry, and polarimetry has been made in order to define the  dynamic, the  chemical constitution in stars, dust and gas, and the origin and evolution of the spectacular spiral pattern. The system was first suspected to be gravitationally interacting by Zwicky (1953), but the confirm was due to the pioneering works of Toomre \& Toomre (1972). The numerical simulations of these authors, starting from an undifferentiated disk, reproduced the formation of outer arms, tails and bridge structures, demonstrating a gravitational interaction with the companion. The system has been investigated in all his components, neutral gas (HI), ionized gas (HII), dust, and stars, and during the last years the studies went on also using balloon borne instrumentation to investigate the far UV, and Hubble Space Telescope to investigate with the highest resolution the fine central spiral structure (Panagia et al. 1995). As a result of this large effort the most part of the characteristics of the system have been explained, but even if we have a general convincing physical model on the origin and dynamic of this structure, more accurate simulations and observations disclosed, as usual, some discrepancies. In fact, the self-gravitating numerical simulations of Toomre (1978) failed to reproduce the inner regions of this system, as well as the efforts by Byrd \& Howard (1989), that obtained a model in which they correctly derived the origin of the two large outer material arms from a gravitational encounter in progress since 70 millions of years, but they were not able to reproduce the inner density wave structure. Observations on inner structure from Zaritsky et al. (1993), using near IR (K band, 2.2$\mu$) photometry in order to overcome the dust and gas reddening and absorption, indicate that the data are consistent with a coherent presence of the two arms winding through about three revolutions toward the galactic centre. According to the authors these results are not consistent with the current theories of the spiral structures. On the other hand, using balloon borne detectors, to detect the far UV ($\sim$2000 \AA), Petit et al. (1996), found different distributions between the evolved, hot, intermediate massive stars (2$\div$5 M$\odot$), and the very massive stars coexisting with the ionized gas (HII regions). This scenario is compatible with the wave density triggered star formation, and then it's in agreement with the theory of the unperturbed spiral structure.
\begin{figure} [h]
\includegraphics{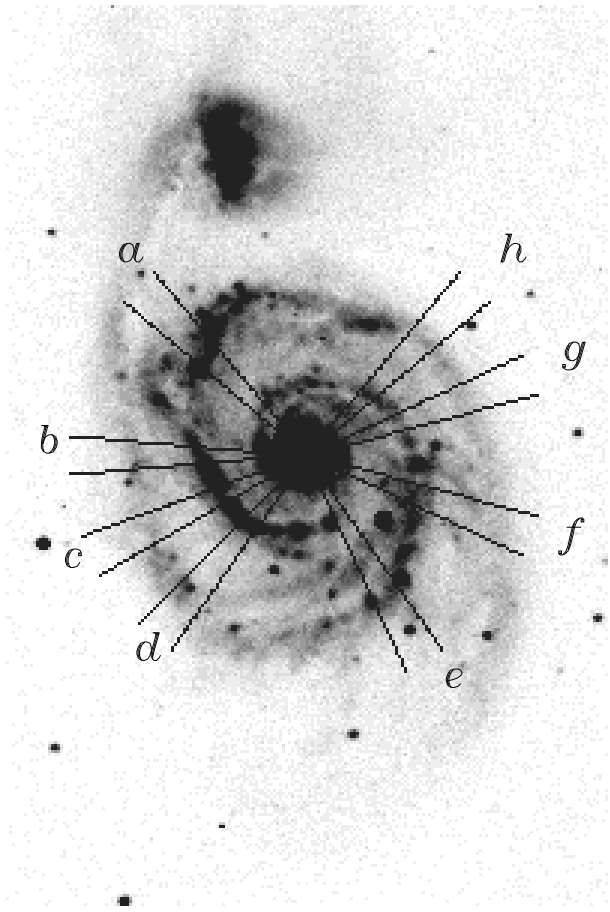}
\caption{\footnotesize The image of M51 in B band showing the sectors from which we derived the radial color indices. The sectors range from the galactic center to 200 arcsec }
\label{Fig.1}
\end{figure}

The hypothesis that density waves and material arms due to tidal interaction can coexist in the same system is already present in a paper of Sundelius (1989), in which, with a self-gravitating numerical simulation, in the late evolution of the interacting system appear both the density waves and the material arms. The idea that in the M51 system the outer structure coming from the gravitational interaction, and the inner structure can be distinct is present also in Elmgreen (1989), and in Howard \& Byrd (1990). The hypothesis of Elmgreen is that the inner density waves were triggered from the tidal waves due to the encounter with the companion. Howard \& Byrd on the contrary, with the results of a numerical self-gravitating simulation suggested that the large outer HI regions detected in the radio from Rots et al. (1990) can be originated by a precedent encounter of M51 with the same companion. Similar conclusions are in Rix \& Rieke (1993) which affirm that some morphological characteristics revealed with photometric observations in the near IR (K band), can be due to the interference between the inner spiral structure and the material tidal arms stimulated by the encounter. In the paper, the authors suppose that the inner structure was pre-existing to the encounter. Having in mind these dilemma, our team attempts to investigate the main arms structures, in order to detect features and/or anomalies in the trend of the density waves which could help in finding some answers to the questions described.

\section{Observations and Measurements}
The results presented in this paper have been carried out from B, V, and I broad bands observations of the M51 system. How we cited in the introduction, several authors used, as star formation tracer, near IR photometric bands, to avoid extinction and reddening due to dust and gas, and to detect a single star population. 
\begin{figure}
\includegraphics{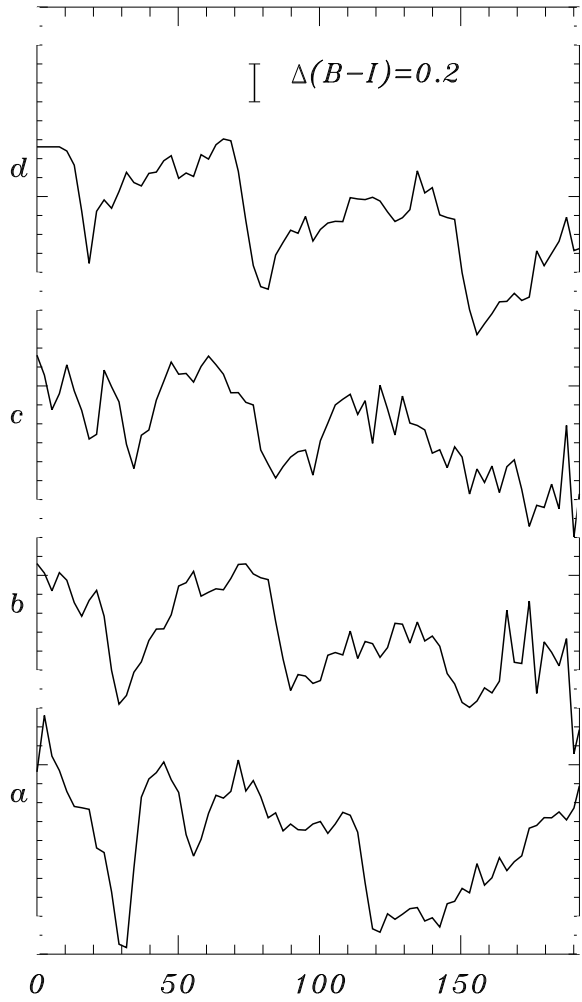}
\caption{\footnotesize Sector B-I radial color index in the EAST part of M51;  The ascissa are in arcsec from the center of the galaxy; a) $\theta=135^0$; b) $\theta=180^0$; c) $\theta=205^0$; d) $\theta=230^0$. The angles are measured starting from the West point counteclockwise}
\label{Fig.2}
\end{figure}

\begin{figure}
\includegraphics{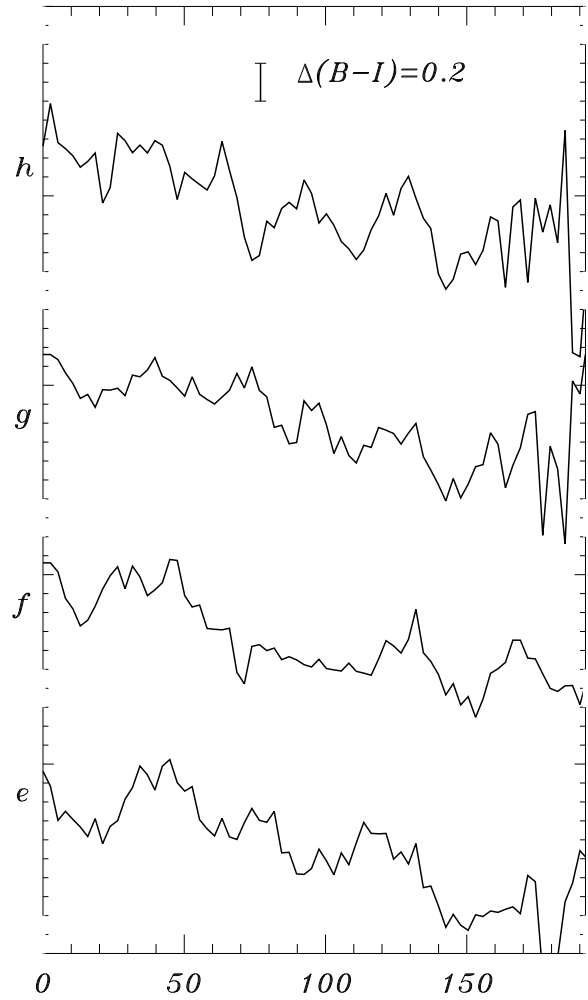}
\caption{Sector B-I radial color index in the WEST part of M51; e) $\theta=300^0$; f) $\theta=340^0$; g) $\theta=20^0$; h) $\theta=45^0$ . The ascissa are in arcsec from the center of the galaxy.}
\label{Fig.3}
\end{figure}

\begin{figure}
\includegraphics{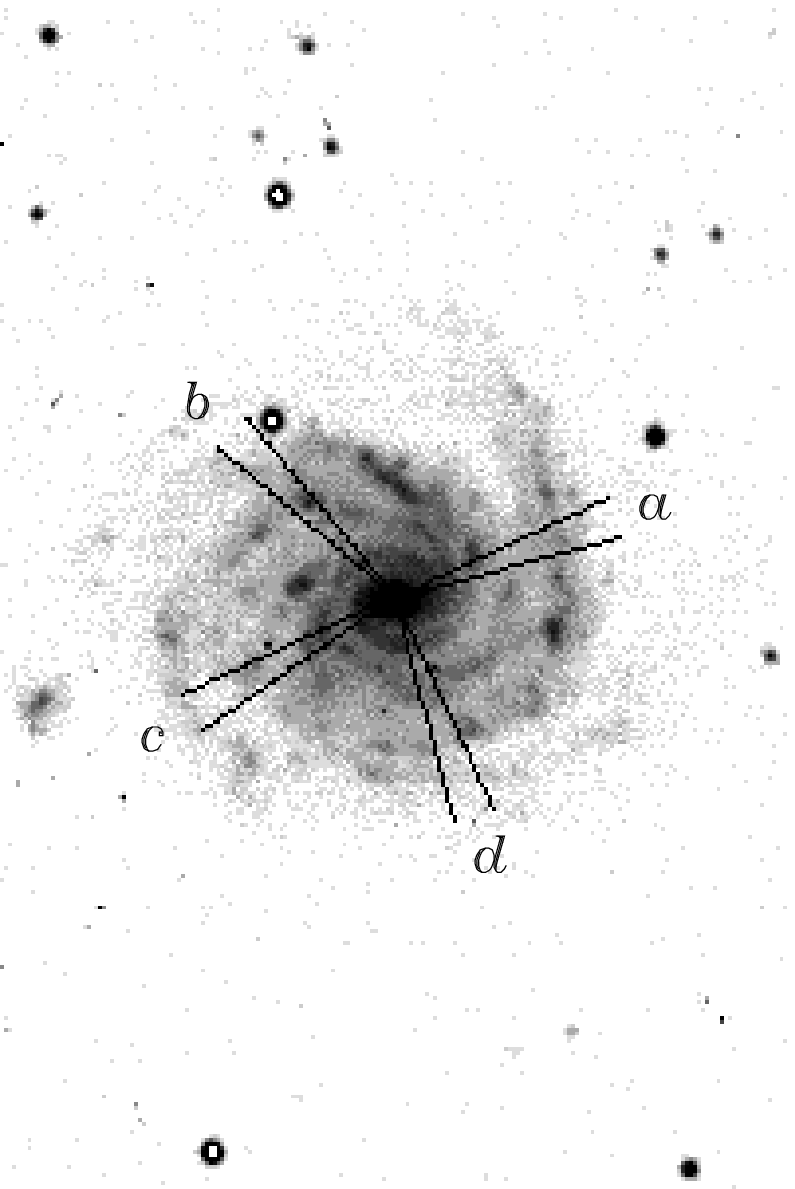}
\caption{The image of NGC 1232 in B band showing the sectors from which we derived the radial color indices. The sectors range from the galactic center to 150 arcsec }
\label{Fig.4}
\end{figure}
\begin{figure}
\includegraphics{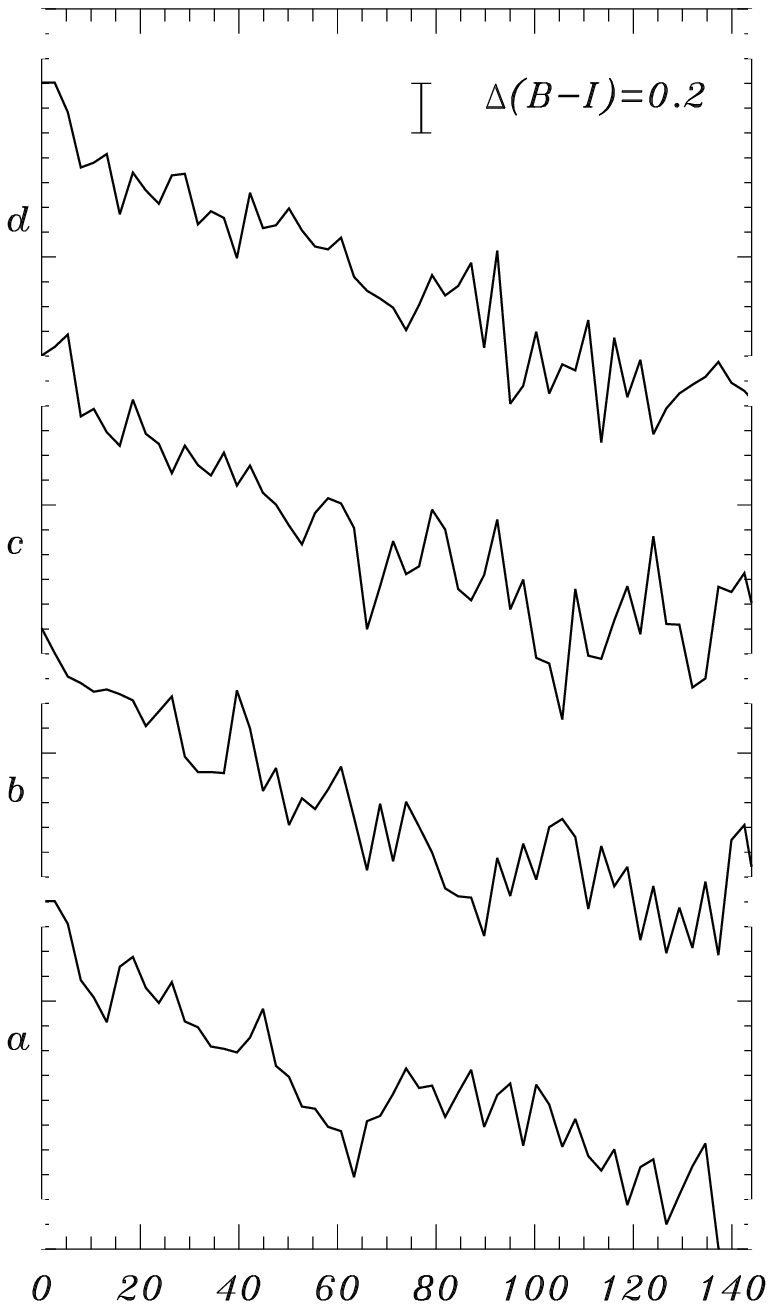}
\caption{Sector B-I radial color index in NGC 1232; a) $\theta=20^0$; b) $\theta=135^0$; c) $\theta=210^0$; d) $\theta=290^0$. The ascissa are
in arcsec from the center of the galaxy.}
\label{Fig.5}
\end{figure}

Since our target was the investigation of the density of the supermassive, just born, short living stars connected with the higher density regions of the galactic materials, across the spiral arms, a color index tracing the temperatures seemed us more fitting the goal. The images were obtained with a CCD Thomson (384x576)x23$\mu$ UV coated at the prime focus of the 60x90x180 cm Schmidt telescope of the Campo Imperatore observatory, since 1990/92, as part of a more general observational program about the radial luminosities and color gradients in face-on nearby spiral galaxies (Pompei \& Natali 1997). 
The M51 system was included in the sample, in order to detect differences with respect to the non-interacting galaxies. Due to the scale factor of the telescope (2.64 arcsec/pixel), the CCD single frame (17x25 arcmin), detected both M51 and companion, including all parts of the outer arms as well as the sky background. In the cited paper the analysis of the images we carried out, was the traditional one used in the studies of radial luminosities and color gradients of galaxies. The {\bf$\mu$(r)} derived from the medians of the luminosities {\bf$\mu$(r,$\theta$)}, were computed over {\bf all the azimuthal angles $\theta$}. The median, as well known, enhances the disk, fading out the characteristics of the spiral arms. The results about the M51 system didn't point out any specific feature with respect to the non-interacting galaxies. Only the radial color trends of this galaxy showed more noise with respect to the signal. In the new measurements, on the contrary, the target being to enhance the spiral arms special features, we tried to use as tracer of the star formation loci, the (B-I) vs r, averaged for {\bf small} $\Delta\theta$ {\bf azimuthal angles} and computed for different {\bf$\theta$}. The preprocessing of the frames i.e. flatting-field, residual gradient subtraction, cosmic rays removing, has been the same as in the previous study. The specific use of the (B-I) color index as tracer, has been suggested from the results of a study about the characteristic of this index, which, close correlated to the temperature of the object like the more used (B-V), presents minor relative errors with respect to the (B-V) itself (Natali el al. 1994). Regarding our target, this tool has been an important factor to reveal features otherwise hidden in the noise. For each photometric band we computed the surface mean (not the median) luminosity {\bf$\mu$(r)} covering only a {\bf sector} of the galaxy with an angle $\Delta\theta=10^0$; fig.1. In fig.2 and fig.3 we report four radial color indices for different azimuthal angles selected  in the EAST part, and in the WEST part respectively of the galaxy. In addition, to have a check, we derived the same sector measurements from NGC 1232 and NGC 3184, two face-on, not-interacting, spiral galaxies. Relative images and radial color indices are in fig.4, fig.5, fig.6 and fig.7.

\section{Results and Discussion}
As discussed in the previous section, we will consider the color indices as density waves tracer for the galactic {\it materia}. The most evident feature  striking in the curves of fig.2 and fig.3, are the different structures of the color indices and then of the density waves in the EAST part of the galaxy, close to the probable orbit of the companion, with respect to the WEST part. In the EAST part, the color indices show in almost all the analyzed azimuthal angles a characteristic saw-toothed structure.
In fact, the trends of the density wave going toward the inner regions rise slownly, (smaller B-I means younger stars, and then higher star formation rate, and then higher material density), dropping after the maxima very quickly. This feature is more evident for the $230^0$ azimuthal angle (fig.1d). For this angle the saw-toothed shape of the density wave is clearly present as far as the inner spiral arm. Moreover, the saw-toothed shape is present also in the outer arm where the density wave, and the SFR, is more faint with respect to the central one. The same characteristics of the radial color index are present also with different angle averages i.e. for {\bf$\Delta\theta = 7^0$}, {\bf$\Delta\theta = 15^0$}, and {\bf$\Delta\theta = 20^0$}. This characteristic is not evident in the diaphram West-East photoelectric scans of M51 reported by Burkhead (1978). About the color index variations, they are stronger in the EAST part, ($\Delta(B-I) = 0.7$), with respect to the WEST part, ($\Delta(B-I) = 0.3$). 
\begin{figure}
\includegraphics{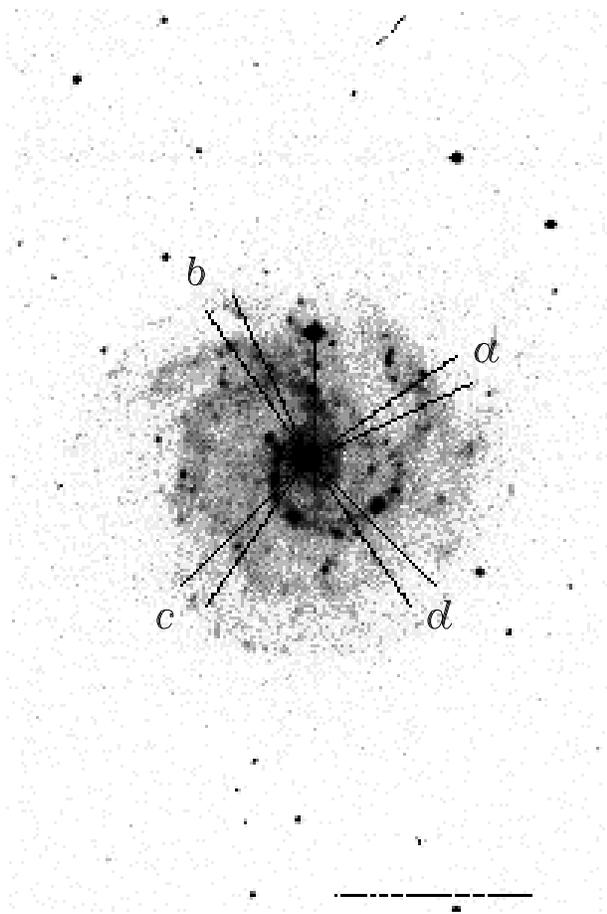}
\caption{The image of NGC 3184 in B band showing the sectors from which we derived the radial color indices. The sectors range from the galactic center to 150 arcsec }
\label{Fig.6}
\end{figure}
\begin{figure}
\includegraphics{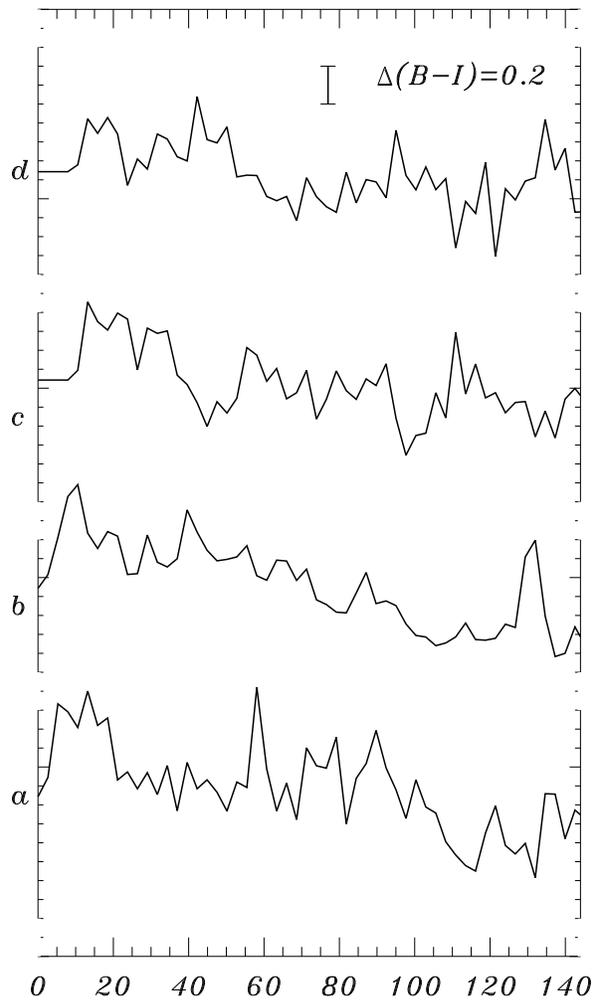}
\caption{B-I radial color index in NGC 3184; a) $\theta=30^0$; b) $\theta=120^0$; c) $\theta=230^0$; d) $\theta=310^0$. The ascissa are in arcsec from the center of the galaxy.}
\label{Fig.7}
\end{figure}
A difference between the EAST and WEST part, about the SFR intensity in M51, was detected with far UV, balloon borne instrumentation early in the 1990 by Bohlin et et al. (1990). In the WEST part, when it is possible to extract the spiral arms signal from the noise, ($\theta=45^0$, or $\theta=300^0$) the radial color index  do not show different rise and fall trends (fig.3). We detected a similar behaviour in the {\it normal} spiral galaxies NGC 1232 and NGC 3184 (fig.4 and fig.6), where the radial color indices present small symmetrical variations in correspondence of the relative spiral arms (fig.5 and fig.7). These results suggest that the saw-toothed morphology we detect in the EAST part of M51 can be due to the tidal forces triggered by the encounter with the companion. The asymmetrical shape of the density wave is consistent with a scenario of a progressive wave propagating in a medium in which the phase velocity is modulated by a pre-existing quasi-steady density wave. In fact, crossing the higher density regions, the front of the progressive wave will decrease his phase velocity generating the observed asymmetries.
%________________________________________________________________

\section{Conclusions}
As reported in the introduction the M51 system presents some characteristics which remain unexplained. Several authors derived from their measurements and/or numerical simulations, the suspect that the companion is responsible only in part of the observed scenario. In many papers it appears the hypothesis that a quasi-steady inner density wave is coexisting with a material tidally induced spiral pattern. The results presented in this paper show a peculiar density shape in the EAST part of M51 clearly connected with the gravitational perturbation of the companion, not revealed in the WEST part of the galaxy or in similar not-perturbed spiral galaxies. The saw-toothed shape of the material arm density in the EAST part of M51, is similar to that of a progressive wave propagating in a medium with a preexisting oscillating density. The variations of the medium density in fact, will produce an oscillating group velocity, generating then, the observed shapes in the progressive wave. In this case it should have to assume the whirlpool was a spiral galaxy before the encounter. We think that our conclusions are not definitive and even if, it is very simple and exciting, it raises new striking questions. In fact, if the galaxy was a spiral before the encounter, the numerical simulations of the system must start from a pre-existing spiral pattern and not from an undifferentiated disk. And then, can the spiral survive to the disrupting tidal forces? If so, which are the new impact parameters as the time-scale of the encounter, the minimum distance of the barycentres, the orbits and the initial velocities and masses of the two objects? What about on other M51-type spirals? Do they present the same kind of density structure in their arms? If so, the progressive wave hypothesis could be strengthened, but only further observational data can solve the new question. 

\hfill

{\em Acknowledgements}. We would like to thank Prof. Nino Panagia for the useful discussions and suggestions about the striking conclusions related to the color gradients presented in this paper. 
%__________________________________________________________________

\end{document}